\documentclass[12pt]{article}

\usepackage[numbers]{natbib}
\usepackage{times}
\usepackage{fancyhdr}
\usepackage{latexsym,amsmath}
\usepackage[pdftex]{graphicx}
\usepackage{multirow}
\usepackage{amssymb}
\usepackage{mathtools}
\usepackage{xcolor}
\usepackage[normalem]{ulem}
\usepackage{url}
\usepackage[colorlinks=true,allcolors=blue]{hyperref}
\usepackage{xurl}
\usepackage{float}
\usepackage{setspace}
\usepackage{booktabs}   
\usepackage{array}      

\title{\vspace*{-1cm}
Topological analysis of hemodynamic response to cardiac resynchronization therapy}

\author
{Aina Ferr\`a Marc\'us$^{1\,\ast}$, 
Carles Casacuberta$^{1}$,
Josep Vives$^{2}$,
Joan Guich$^{1}$,\\
Gerard Amor\'os-Figueras$^{3\,\dagger}$,
Jose M. Guerra$^{3\,\dagger}$
\\[0.3cm]
\normalsize{$^{1}$Department of Mathematics and Computer Science,}\\
\normalsize{Universitat de Barcelona, Barcelona, Spain}\\
\normalsize{$^{2}$Department of Economic, Financial and Actuarial Mathematics,}\\
\normalsize{Universitat de Barcelona, Barcelona, Spain} \\
\normalsize{$^{3}$Department of Cardiology, Hospital de la Santa Creu i Sant Pau},\\ 
\normalsize{Institut de Recerca Sant Pau (IR Sant Pau),}\\ 
\normalsize{CIBERCV, Universitat Aut\`onoma de Barcelona, Barcelona, Spain} \\[0.4cm]
\normalsize{$\dagger$ These authors share senior authorship}
\\[0.4cm]
\normalsize{$^\ast$Corresponding author. E-mail: aina.ferra.marcus@gmail.com}
}

\date{}

\begin{document}

\maketitle 

\begin{abstract}{\noindent
\underline{Objective:}
The Mapper algorithm is a qualitative method in topological data analysis that constructs graphs from point clouds by combining dimensionality reduction and clustering techniques. The aim of this study is to apply Mapper, together with novel quantitative indices, to compare the effects of biventricular pacing from the left ventricular epicardium versus the endocardium in a swine model of pacing-induced non-ischemic cardiomyopathy.
\\[0.2cm] \underline{Methods:}
The distributions of four hemodynamic variables from a previous study on endocardial and epicardial cardiac resynchronization in an experimental swine model of non-ischemic cardiomyopathy were analyzed using the Mapper algorithm, enhanced with numerical indices quantifying self-connectivity, scattering, and homogeneity of the resulting colored graphs. \\[0.2cm]\underline{Results:} 
Statistically significant differences were observed between pacing from basal regions versus mid or apical regions, with the following self-connectivity index values: basal $0.57$; mid $0.14$ ($p<0.01$); apical $0.24$ ($p<0.01$). Endocardial stimulation at lateral sites increased the contrast between the distributions of basal versus mid or apical data, when compared with epicardial stimulation. \\[0.2cm]
\underline{Conclusions:} Topological analysis using the Mapper algorithm, enhanced with quantitative statistical measures, revealed new and biologically plausible significant differences in pacing effects across heart regions.}
\end{abstract}

\section{Introduction}

Topological methods for data analysis have been widely applied across a variety of disciplines since the decade of 2010. While numerical descriptors based on persistent homology are the most frequently used~\cite{Carlsson2009}, the Mapper algorithm is a qualitative method in topological data analysis, introduced in~\cite{Mapper}, that transforms point clouds into graphs by combining dimensionality reduction and clustering techniques.
The Mapper algorithm has been applied in fields such as industry, biomedicine, finance, and many others \cite{Dlotko,Lum,Nicolau}. It~has also been used in artificial intelligence to analyze the structure of neural network activations and to assess their generalization capacity~\cite{purvine}.
Mapper is most often used in the context of unsupervised learning ---common applications include visualization, clustering, and feature discovery 
\cite{Hinks, Rucco, Yao}.
It has proved particularly useful in medical studies, owing to its ability to capture the global structure of datasets even when sample sizes are limited. A~further advantage of Mapper is its interpretability: it produces graphs that highlight and summarize the relevant features of the data in a visually accessible way.

Even though Mapper has traditionally been applied as a qualitative exploratory tool, relying on visualization of the resulting graphs, there has been significant progress toward defining quantitative measures for analyzing its
resulting graphs. In~\cite{Li}, a patient-patient similarity network was constructed using Mapper and applied enrichment analysis to statistically assess the overrepresentation of
type~2 diabetes within nodes, as well as to identify comorbidity patterns. In~\cite{Liao}, spatial analysis of functional enrichment was used to compute enrichment scores across nodes and their neighborhoods, thus highlighting localized associations in microbiome data. More generally, a mathematical framework was proposed in \cite{Kalyanaraman}
for quantifying interesting paths in Mapper graphs, providing optimization formulations to systematically identify substructures of interest. The enhancement of Mapper graphs with statistical methods has also been explored in~\cite{Tao-Ge}.

In this article, we introduce a quantitative procedure aimed at obtaining statistically significant results using Mapper, applied to data from a cardiac resynchronization study reported in~\cite{Europace,Soriano_etal}.
While the use of quantitative indices from graph theory, such as homophily indicators~\cite{kim-altmann}, is well established in network science, the use of numerical descriptors for Mapper graphs constitutes a novel methodological contribution within topological data analysis.  

The main objective of the study reported in \cite{Europace} was to compare the effects of biventricular pacing 
from left ventricular epicardium versus the endocardium in a swine model of pacing-induced non-ischemic cardiomyopathy.
The results presented here are biologically plausible and consistent with those obtained in~\cite{Europace}, achieving greater statistical power. 
Thus, our work provides a 3R enhancement (Replacement, Reduction, and Refinement), in the framework of animal research, of the experimental approach used in~\cite{Europace}.
Specifically, integration of several configurations of the cardiac resynchronization device yields an expanded dataset, 
facilitating the attainment of statistical significance through our topology-based method.

\section{Materials and methods}

\subsection{Dataset}
\label{dataset}

This study was performed using data from~\cite{Europace}, where six female domestic swine with pacing-induced non-ischemic cardiomyopathy were submitted to endocardial or epicardial biventricular pacing. Different electrodes were used to pace the right atrium, the right ventricle, and the left ventricle.
With the aim of comparing the effect of the pacing site, electrodes were placed at three basal regions (anterior, lateral and posterior), three mid regions (anterior, lateral and posterior) and two apical regions (anterior and lateral), both from the epicardium and from the endocardium. In addition, at each site, six different configurations of the resynchronization therapy (CRT) device were performed (two atrioventricular delays: $80$ and $110$~ms  and three interventricular intervals: $0$, $30$, and $-30$~ms).
Electrocardiograms and intra-cavitary pressure probes were registered and signals were digitized for subsequent analysis. Values obtained after biventricular pacing were compared with baseline values obtained during previous dyssynchronous pacing, and differences were expressed as percentages of change.

The hemodynamic variables 
in this study are left-ventricular peak pressure (LVP), the maximum and minimum time derivatives of LVP (DPDT+ and DPDT--), and mean aortic blood flow (ABF). However, the minimum derivative of LVP (denoted DPDT--) reflects different physiological features compared to the other three variables. For this reason, we repeated our analyses both including and excluding DPDT--, and compared the corresponding self-connectivity indices of Mapper graphs. We also assessed the effect of potential outliers by recalculating our results while excluding one individual at a time.

Thus, our study dataset is a point cloud consisting of $576$ points in a $4$-dimensional space, whose coordinates are percentages of change of the four study variables with respect to their baseline values. For each of the six study individuals (labelled as FIS06, FIS13, FIS14, FIS15, FIS18, and FIS19), the point cloud contains $36$ basal points, $36$ mid points and $24$ apical points, half of which are endocardial and the rest are epicardial, obtained with six CRT device configurations. 

The study protocol was reviewed and approved by the Institutional Animal Care and Use Committee of the Institution and the Animal Experimental Committee of the local government (Generalitat de Catalunya, authorization {\#}11472) in accordance with Spanish Law (RD 53/2013) and European Directive 2010/63/EU.

\subsection{Topological data analysis}
\label{TDAmapper}

The Mapper algorithm was introduced in \cite{Mapper} for qualitative analyses based on shape features of point clouds. Mapper is a combination of dimensionality reduction and clustering techniques, yielding graph representations of dataset structures.
Its main ingredients are a filter function and a clustering procedure. 
The filter function $f$ is defined on the dataset and takes values in the real numbers $\mathbb{R}$ or in a multidimensional space $\mathbb{R}^d$. The image of $f$ is covered with a finite number of overlapping open sets $\{U_i\mid i\in I\}$, where the cardinality of $I$ and the size of the covering sets $U_i$ are parameters to be chosen in the algorithm. For each $i\in I$, the subset $f^{-1}(U_i)$ is split into clusters $C_{i_1},\dots,C_{i_{n_i}}$. The output of the algorithm is an undirected graph whose set of nodes is in bijective correspondence with the set of clusters and there is an edge between nodes $i_r$ and $j_s$ if the subsets $C_{i_r}$ and $C_{j_s}$ have nonempty intersection.
Hence the nodes of the graph are weighted by the number of data points contained in the corresponding cluster.
In this study, we have chosen principal component analysis (PCA) as filter function ---by projecting the data point cloud onto a variance-maximizing $2$-dimensional parameter space--- and we have used $k$-means as clustering algorithm. Since $k$-means includes a random seed, each instance of Mapper was run five times and the resulting quantitative summaries were averaged. When $p$-values are indicated, these are the median values of the five repetitions.

It should be emphasized that Mapper graphs are purely relational structures. Thus, while adjacency is crucial, distances between nodes are meaningless and location of nodes in graph pictures is irrelevant. In this paper, Fig.\,\ref{fig:1} and Fig.\,\ref{fig:2}, as well as those in the Supplementary Material, have been produced using the Python library NetworkX \cite{networkx}.

\begin{figure}[htbp]
    \centering    \includegraphics[width=0.6\textwidth]{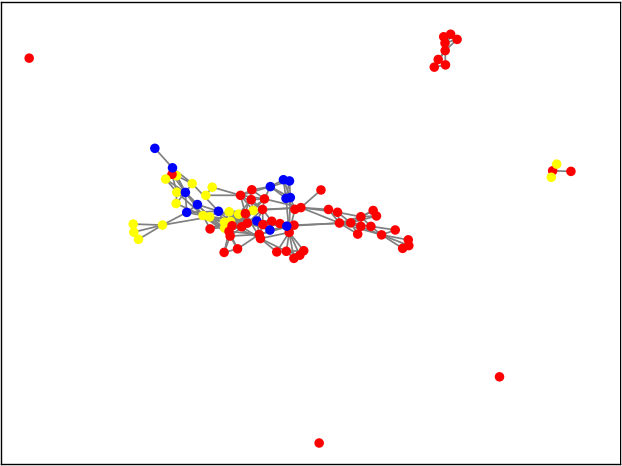}
    \caption{Colored Mapper graph of the study dataset for a comparison between heart sites, obtained with percentages of change of four hemodynamic variables (LVP, DPDT+, DPDT--, ABF). Each node corresponds to a cluster of points in $4$-dimensional space, and each point corresponds to one of six study individuals paced at a heart site (basal, mid, or apical), at endocardial or epicardial location, and using one of six CRT device configurations. Each node is colored with the dominant color among the points in the corresponding cluster.
    Colors of nodes correspond to heart sites (red: basal; yellow: mid; blue: apical).}
    \label{fig:1}
\end{figure}

Although Mapper is a descriptive tool that yields graphs as outcomes, in this study we have defined three indices for quantitative analysis of colored Mapper graphs, as described in Subsection~\ref{descriptors}.
We suppose that points in the dataset are labelled with a set of colors ---basal points are red, mid points are yellow, and apical points are blue--- and have customized Mapper (retrieved from \cite{KeplerMapper}) so that nodes of graphs are also labelled with the same set of colors. Since each node of the graph corresponds to a cluster of points in the dataset, we assign to a node the dominant color among the points in the corresponding cluster (Fig.\,\ref{fig:1}). In case of ties, the dominant color among adjacent nodes is chosen, and if this does not break a tie then the dominant color in the graph is selected.

\subsubsection{Numerical descriptors}
\label{descriptors}

Let $G=(V,E)$ be a colored Mapper graph with colors $\{c_1,\dots,c_k\}$, where $V$ is its set of nodes (or vertices) and $E$ is its set of edges. Then
$V=V(c_1)\cup\cdots\cup V(c_k)$,
where $V(c_i)$ denotes the set of nodes of color~$c_i$. A~node $v\in V(c_i)$ may contain data points of several colors, with a majority of $c_i$-colored points. The number of elements of $V(c_i)$ is denoted by~$N_i$ in what follows.

\vskip 0.3cm

\noindent
\textbf{Self-connectivity.}
A node $v\in V(c_i)$ is called \emph{insular} if there is no edge from $v$ to a node $w\in V(c_j)$ with $c_j\ne c_i$. Nodes with no incident edges are also considered insular.
For each $i\in\{1,\dots,k\}$, let $r_i$ be the number of insular nodes of color $c_i$.
The \emph{self-connectivity index} of color $c_i$ is defined as the ratio $r_i/N_i$ between the number of $c_i$-colored insular nodes and the total number of $c_i$-colored nodes:
\begin{equation}
\label{SC}
{\rm SC}_i(G)=\frac{r_i}{N_i}=
\frac{\#\{\text{$c_i$-colored nodes connected with $c_i$-colored nodes only}\}}{\#\{\text{$c_i$-colored nodes}\}}.
\end{equation}
This is a proportion, therefore suitable for statistical comparison tests. We compare proportions $p_1$ and $p_2$ from sample sizes $n_1$ and $n_2$ using the estimator
\[
\frac{p_1-p_2}{\sqrt{p_c(1-p_c)(1/n_1+1/n_2)}}
\]
for a two-tailed standard normal test, where $p_c=(n_1p_1+n_2p_2)/(n_1+n_2)$.

In graph theory there is a similar measure, called \emph{homophily index}, which is a ratio of the probability of linkages between nodes belonging to different groups by the probability of linkages between nodes within the same group~\cite{kim-altmann}. Network homophily refers to the phenomenon that nodes tend to connect with similar nodes; it has alternatively been defined as the fraction of edges that connect nodes with the same labels~\cite{ma}.

\vskip 0.3cm

\noindent\textbf{Homogeneity.} For each node $v\in V(c_i)$ and $j=\{1,\dots,k\}$, let $m_{j}$ be the number of $c_j$-colored data points within~$v$. The ratio
\[
h_i(v)=\frac{m_i}{m_1+\cdots+m_k}
\]
is the proportion of $c_i$-colored points within the node~$v$, where $k$ is the number of colors.
The \emph{homogeneity index} of a colored Mapper graph $G=(V,E)$ for color $c_i$, which we denote by $h_i(G)$, is the average value of $h_i(v)$ over all nodes $v\in V(c_i)$.
Hence, for each color~$c_i$, the homogeneity index is the average percentage of $c_i$-colored data points within $c_i$-colored nodes. This indicator is used to detect a possible bias due to an unbalanced distribution of colors within the nodes of the graph. 
In the context of neural network analysis, there is a similar notion called \emph{node-wise purity}, which is defined for each node $i$ as $1/c_i$, where $c_i$ is the number of class labels in node~$i$; see \cite{purvine}.

\vskip 0.3cm

\noindent\textbf{Scattering.} 
For each color $c_i$, the \emph{scattering index} of a colored Mapper graph $G=(V,E)$ is defined as
\[
s_i(G)=
\frac{N_i/(N_1+\cdots+N_k)}{D_i/(D_1+\cdots+D_k)}
\]
where $D_i$ is the number of $c_i$-colored data points. For each color~$c_i$, the scattering index is the ratio between the number of $c_i$-colored nodes and the total number of nodes, divided by the ratio between the number of $c_i$-colored points in the dataset and the total number of points in the dataset: 
\[
s_i(G)=
\frac{(\#\{\text{$c_i$-colored nodes}\})/(\#\{\text{nodes}\})}{(\#\{\text{$c_i$-colored data points}\})/(\#\{\text{data points\})}}.
\]
This ratio measures the dominance of color $c_i$ in the collection of clusters obtained within the Mapper algorithm.
Hence, a high scattering index value (greater than $1$) for color $c_i$ indicates that the proportion of $c_i$-colored graph nodes among all nodes is higher than expected based on the distribution of color $c_i$ in the original point cloud. The most plausible interpretation is that $c_i$-colored data points tend to be more scattered throughout the dataset, thus forming smaller clusters, compared to other colors.

\section{Results}
\subsection{Comparison between left ventricle locations}

Fig.\,\ref{fig:1} shows the joint distribution of percentage changes of four 
hemodynamic variables.
Graph nodes correspond to clusters of points in $4$-dimensional space (LVP, DPDT+, DPDT--, ABF), and each node is colored with the dominant color among the points in the corresponding cluster. Graph edges indicate overlapping between neighboring clusters. Red-colored nodes, corresponding to study individuals paced at the basal site, are more scattered than yellow-colored nodes (mid site) and blue-colored (apical site) nodes. Moreover, the self-connectivity index is higher for red-colored nodes, indicating a higher proportion of insular nodes with respect to the total number of red-colored nodes.

Values of self-connectivity, scattering and homogeneity indices obtained from the graph in Fig.\,\ref{fig:1} are shown in Table~\ref{table:BMA}, where self-connectivity and scattering are seen to be higher for the basal site, while node homogeneity is similar for all sites.
Corresponding $p$-values of comparison tests for self-connectivity indices are shown in~Table~\ref{table:BA-BM-AM}.

\begin{table}[htbp]
\centering
\begin{tabular}{rccc}
 & \textbf{Self-connectivity} & \textbf{Scattering} & \textbf{Homogeneity} \\ \hline \\[-0.25cm]
\textbf{Basal}& 0.57 & 1.67 & 0.78 \\[0.12cm]
\textbf{Mid} & 0.14 & 0.57 & 0.69 \\[0.12cm]
\textbf{Apical} & 0.24 & 0.61 & 0.77  \\[0.05cm] \hline \end{tabular}
\caption{Quantitative indices of the colored Mapper graph shown in Fig.\,\ref{fig:1} averaged from five repetitions with different seeds. Red: basal; yellow: mid; blue: apical.
Variables: LVP, DPDT+, DPDT--, ABF.}
\label{table:BMA}
\end{table}

\begin{table}[htbp]
\centering
\begin{tabular}{lccc}
& \textbf{Basal vs apical} & \;\textbf{Basal vs mid} & \textbf{Apical vs mid} \\ \hline \\[-0.25cm]
\textbf{\boldmath$p$-value} & $0.0011^{**}$ & \;$ 0.0021^{**}$ & $0.1589$
\\[0.1cm] \hline
\end{tabular}
\caption{Significance values of comparison tests for self-connectivity indices in Table~\ref{table:BMA}. Medians of $p$-values after five repetitions are shown (*: $p<0.05$; **: $p<0.01$; ***: $p<0.001$).
Variables: LVP, DPDT+, DPDT--, ABF.}
\label{table:BA-BM-AM}
\end{table}

The $p$-values in Table~\ref{table:BA-BM-AM} indicate a significant distinction between pacing at basal locations versus mid or apical locations, with a higher self-connectivity index value in case of the basal location. The scattering index is also much higher for the basal region, suggesting that basal points are more dispersed in the original dataset, while homogeneity values indicate that there is no relevant bias attributable to an
uneven distribution of colors across graph nodes.

Since the minimum rising rate of left-ventricular peak pressure (DPDT--) reflects different physiological features than the other study variables, we repeated all calculations using only LVP, DPDT+ and ABF, for the sake of comparison. The results are shown in Tables \ref{table:BMA_APP} and \ref{table:BA-BM-AM_APP} in the Supplementary Material, and lead to similar conclusions as Tables \ref{table:BMA} and~\ref{table:BA-BM-AM}.

\subsection{Endocardial location versus epicardial location}

\begin{figure}[htbp]
    \centering
\includegraphics[width=0.6\textwidth]{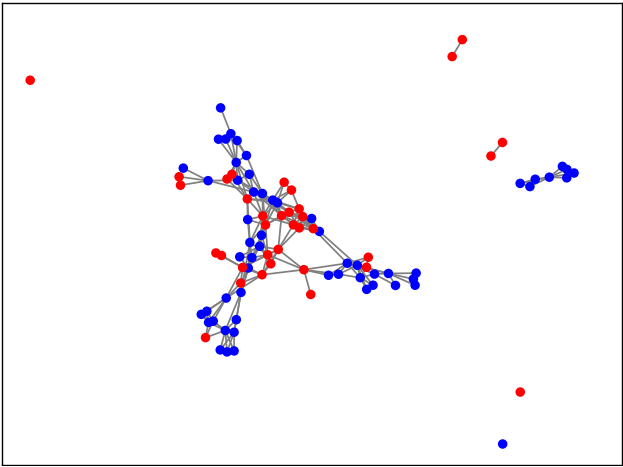}
    \caption{Colored Mapper graph of the study dataset for a comparison between epicardial stimulation versus endocardial stimulation, obtained with percentages of change of four hemodynamic variables (LVP, DPDT+, DPDT--, ABF). Each node corresponds to a cluster of points in $4$-dimensional space, and each point corresponds to one of six study individuals paced at a heart site (basal, mid, or apical), at endocardial or epicardial location, and using one of six CRT device configurations. Each node is colored with the dominant color among the points in the corresponding cluster.
    Red: epicardial; blue: endocardial.}
    \label{fig:2}
\end{figure}

The effect of endocardial pacing was compared with the effect of epicardial pacing by drawing new Mapper graphs and computing the corresponding numerical indices. Although the dataset is the same, there are only two colors now (red: epicardial, blue: endocardial), and therefore the coloring of nodes in the resulting graphs is different, as shown in Fig.\,\ref{fig:2}. 
In this graph, self-connectivity is higher in blue-colored nodes, suggesting an increased departure from average values in the case of endocardial pacing.

\newpage

A significant difference ($0.48$ versus $0.23$, $p<0.01$) was found in a comparison test for self-connectivity indices in endocardial sites versus epicardial sites. Scattering is also higher for endocardial sites ($1.25$ versus $0.75$), while homogeneity is similar in both sites.
The corresponding summaries and $p$-values of comparison of self-connectivity indices are given in Tables \ref{table:EE} and~\ref{table:EEP}. 

\begin{table}[htbp]
\centering
\begin{tabular}{rccc}
 & \textbf{Self-connectivity} & \textbf{Scattering} & \textbf{Homogeneity} \\ \hline \\[-0.25cm]
\textbf{Endocardial} & 0.48  & 1.25   & 0.77                 \\[0.12cm]
\textbf{Epicardial} & 0.23 & 0.75   & 0.74  \\[0.05cm] \hline
\end{tabular}
\caption{Quantitative indices of the colored Mapper graph shown in Fig.\,\ref{fig:2}  averaged from five repetitions with different seeds.
Variables: LVP, DPDT+, DPDT--, ABF.}
\label{table:EE}
\end{table}

\begin{table}[htbp]
\centering
\begin{tabular}{lc}
& \textbf{Endocardial vs epicardial} \\ \hline \\[-0.25cm]
\textbf{\boldmath$p$-value} & $0.0032^{**}$ \\[0.1cm] \hline
\end{tabular}
\caption{Significance $p$-value of a comparison test for self-connectivity indices in Table~\ref{table:EE}. Median of $p$-values after five repetitions is given.
Variables: LVP, DPDT+, DPDT--, ABF.}
\label{table:EEP}
\end{table}

In order to assess the importance of anterior regions versus lateral or posterior regions at each location in the distinction between endocardial and epicardial sites, self-connectivity indices were calculated by grouping regions (anterior, lateral and posterior) in all possible combinations. This was first done for all locations together (basal, mid and apical), and then grouping only basal and mid zones to estimate the relative contribution of the apical zone. Corresponding $p$-values are shown in Table~\ref{table:ALP}. 
These results suggest that the lateral region consistently discriminates between endocardial and epicardial pacing, while the posterior region shows lowest discrimination capacity. 

\begin{table}[htbp]
\centering
\begin{tabular}{rlllllll}
 & \textbf{ALP} & \textbf{AL} & \textbf{AP} & \textbf{LP} & \textbf{A} & 
 \textbf{L} &
 \textbf{P} \\ \hline \\[-0.2cm]
 \textbf{With apical}\;\;\; &
 $0.0139^*$ & $0.0039^{**}$ & $0.0260^*$ & $0.3535$ &  $0.3051$ & $0.0003^{***}$ &  $0.2106$ \\[0.12cm] \textbf{Without apical}\;\;\; &
 $0.1957$ & $0.1589$ & $0.2436$ & $0.1306$ & $0.4041$ & $0.0638$ & $0.3290$ \\[0.05cm] \hline
\end{tabular}
\caption{Comparison between endocardial and epicardial sites at each combination of regions (A: anterior; L: lateral; P: posterior). Medians of $p$-values are shown for self-connectivity indices after five repetitions of the Mapper graphing (*: $p<0.05$; **: $p<0.01$; ***: $p<0.001$).
Variables: LVP, DPDT+, DPDT--, ABF. First row: including apical data in all columns; second row: excluding apical data.}
\label{table:ALP}
\end{table}

Seeking for further evidence of the distinction between endocardial pacing and epicardial pacing, Mapper graphs were generated separately for each subsample (endocardial and epicardial) of the study dataset. The Mapper graph corresponding to epicardial points is shown in Fig.\,\ref{fig:3} (Supplementary Material),
and the one corresponding to endocardial points is shown in Fig.\,\ref{fig:4}.
The results of comparison of self-connectivity indices of basal versus mid and basal versus apical are shown in Table~\ref{table:BMBAEpiEndo} for each of the two graphs (Figs.\,\ref{fig:3} and \ref{fig:4}).
Statistical significance is consistently higher in the case of endocardial pacing.

\begin{table}[htbp]
\centering
\begin{tabular}{rll}
& \textbf{Epicardial} & \textbf{Endocardial} \\ \hline \\[-0.25cm]
\textbf{Basal vs mid}    & \;\;$0.0937$             & \;\;$0.0000^{***}$              \\[0.12cm]
\textbf{Basal vs apical} & \;\;$0.4148$             & \;\;$0.0279^*$     \\[0.05cm] \hline \end{tabular}
\caption{Significance $p$-values of comparison tests for self-connectivity indices of heart sites separated into epicardial and endocardial stimulation.
Variables: LVP, DPDT+, DPDT--, ABF.}
\label{table:BMBAEpiEndo}
\end{table}

Overall, our results support the claim that the self-connectivity index of colored Mapper graphs can be a useful numerical indicator for statistical comparisons. 

\subsection{Effect of outliers}

Outlier individuals could potentially be influencing the above results. To account for this effect, $p$-values were computed anew by excluding each individual, one at a time. Results are shown in Table~\ref{table:excl} of the Supplementary Material and compared with the corresponding results by excluding minimum rising rate of left-ventricular peak pressure (DPDT--) from the analyses 
(Table~\ref{table:excl_noDPDT}).

When taking into account intra-individual variability, it should be reported that one individual, labelled as FIS14, was treated with norepinephrine during the original study~\cite{Europace}. However, no anomalies were detected involving FIS14. While the results in Tables \ref{table:excl} and \ref{table:excl_noDPDT} tend to be consistent, variability in statistical significance arises from the inherent randomness of the Mapper algorithm and the small sample size.

\section{Discussion}

This work demonstrates that
the distribution of hemodynamic variables within basal points in the dataset is significantly different from the distribution within mid points and apical points. The distinction is increased in the case of endocardial stimulation. 

The self-connectivity index and the scattering index introduced in this study are novel contributions and have served as key descriptors in obtaining the results presented in this article.
Higher self-connectivity values are interpreted as indicating singular behavior of the corresponding heart region (basal, mid, or apical), and higher scattering values reflect greater dispersion of the associated data points relative to those from other regions.
Our results indicate that self-connectivity indices of epicardial and endocardial Mapper nodes are also significantly different.

Since Mapper is a descriptive tool for datasets, our conclusions primarily highlight distinctions. What can be asserted from our results is that the response to pacing is significantly dependent on the stimulation site, with endocardial pacing having a greater impact overall. Indeed, as shown in~\cite[Table~3]{Europace}, increases in DPDT+ values are greater when pacing is applied at basal sites, which is associated with more favorable hemodynamic responses (see~\cite[Table~2]{Europace}).
The observation that basal pacing may increase the likelihood of a positive response to cardiac resynchronization therapy more effectively than mid or apical pacing has also been reported in other studies~\cite{Bordachar,Dzemali}.

In~\cite[Table~3]{Europace}, it is also shown that endocardial pacing tends to produce a stronger effect on DPDT+ in basal regions compared to epicardial pacing ($16.2\% \pm 4.1$ vs. $11.9\% \pm 2.1$), although the differences are not conclusive.
Our finding that the contrast between endocardial and epicardial stimulation is mostly localized in lateral or anterior regions could explain why the difference was not found to be statistically significant in~\cite{Europace}.

Further numerical analyses should be undertaken using quantitatively enriched Mapper graphing techniques, aiming for greater overall consistency, possibly by selecting different hemodynamic variables or adjusting the parameters of the Mapper algorithm. While principal component analysis (PCA) has proved to be a convenient filter function, and the choice of clustering algorithm is unlikely to have a major impact, the resulting Mapper graph is sensitive to parameters such as the number of points per cluster, the partitioning of the filter range, and the degree of overlap between intervals.
Since Mapper graphs also depend on a random seed used in the $k$-means clustering algorithm, the $p$-values reported in this article exhibit some variability when computations are repeated on the same dataset. To improve robustness, all Mapper graphs were generated five times, and the median $p$-value of the corresponding comparison tests for the self-connectivity indices was used. The use of medians, rather than averages, reflects more accurately the significance level achieved in each experiment.

\section*{Conflict of Interest Statement}

The authors declare that the research was conducted in the absence of any commercial or financial relationships that could be construed as a potential conflict of interest.

\section*{Author Contributions}
AFM: Writing -- original draft, Formal analysis, Software, Methodology, Validation;
CC: Writing -- original draft, Methodology, Supervision;
JV: Writing -- review \& editing,  Methodology;
JG: Writing -- review \& editing, Formal analysis, Software, Methodology;
GAF: Writing -- review \& editing, Investigation, Data curation, Methodology; JMG: Writing -- review \& editing, Investigation, Methodology, Resources, Funding acquisition.

\section*{Funding}

AFM was supported by MCIN/AEI under doctoral grant PRE2020-094372; AFM and CC were supported by MCIN/AEI through research projects PID2020-117971GB-C22 and PID2022-136436NB-I00; GAF and JMG were supported by grants from the Spanish Ministerio de Econo\-m\'\i a y Competitividad, Instituto de Salud Carlos III (FIS-PI10/01149, PI20/01702) and Fondo Europeo de Desarrollo Regional (FEDER).

\section*{Acknowledgments}
We are grateful to Dr.\ M.\ Teresa Puig for initiating this collaborative study and for her contribution to several team meetings.

\section*{Data and Code Availability Statement}

The datasets analyzed for this  study were obtained from previous studies~\cite{Europace,Soriano_etal}. Code developed for this project is available at~\url{https://github.com/ainafem/mapper_cardiac_analysis}.

\vfill
\eject

\section*{Supplementary Tables and Figures}

The following four variables have been used in the present study: left-ventricular peak pressure (LVP), maximum rising rate of LVP (DPDT+), minimum rising rate of LVP (DPDT--), and mean aortic blood flow (ABF). Since DPDT-- reflects different physiological features compared to the other three variables, we include in this Supplementary Material results of analyses that exclude DPDT--
(Tables \ref{table:BMA_APP}--\ref{table:BMBAEpiEndo_APP}).

\begin{figure}[htbp]
    \centering
\includegraphics[width=0.6\textwidth]{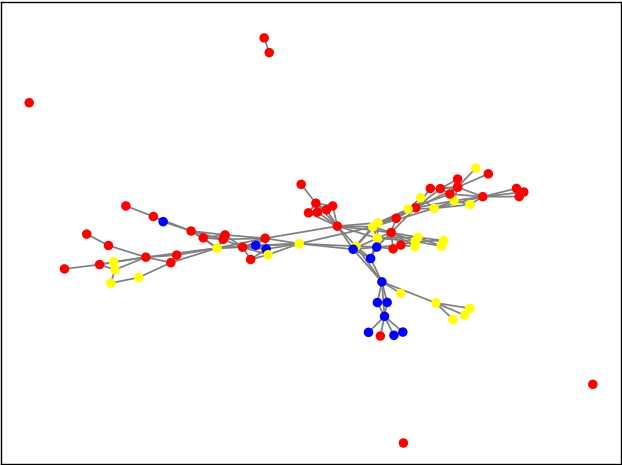}
    \caption{Colored Mapper graph using only \textbf{epicardial} points in the study dataset. Colors of vertices correspond to heart sites (red: basal; yellow: mid; blue: apical).}
    \label{fig:3}
\end{figure}

\begin{figure}[htbp]
    \centering
\includegraphics[width=0.6\textwidth]{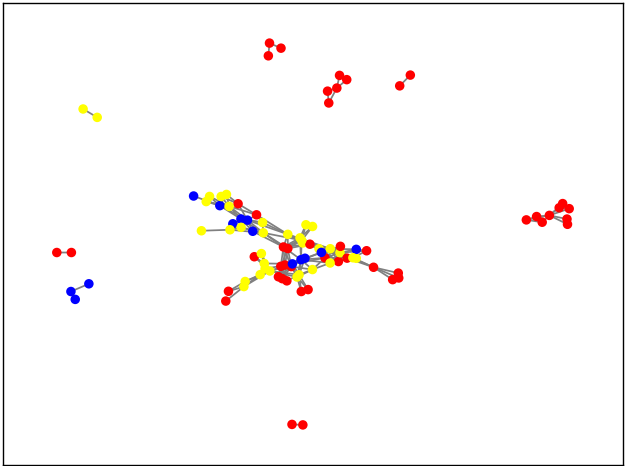}
    \caption{Colored Mapper graph using only \textbf{endocardial} points in the study dataset. Colors of vertices correspond to heart sites (red: basal; yellow: mid; blue: apical).}
    \label{fig:4}
\end{figure}

Our dataset consists of $576$ points, whose coordinates are percentages of change of the four study variables with respect to their baseline values (or three variables when DPDT-- is excluded). For each of the six study individuals, the point cloud contains $36$ basal points, $36$ mid points and $24$ apical points, half of which are endocardial and the rest are epicardial, obtained with six different configurations of the resynchronization therapy device. In this Supplementary Material, Fig.\,\ref{fig:3} refers only to epicardial points, while Fig.\,\ref{fig:4} refers to endocardial points.

Colored Mapper graphs have been calculated with the following parameters:
\begin{itemize}
\item
Lens (filter function): Principal Component Analysis. 
\item
Cover: 7 cubes and 40\% overlapping. 
\item
Clusterer: $k$-means with three clusters.
\end{itemize}

In order to assess the effect of potential outliers, our results were recalculated by excluding one individual at a time. Study individuals are labelled as FIS06, FIS13, FIS14, FIS15, FIS18, and FIS19. The resulting significance $p$-values for self-connectivity index differences between endocardial and epicardial sites are shown in Table~\ref{table:excl} (including DPDT--) and in Table~\ref{table:excl_noDPDT} (excluding DPDT--).

\begin{table}[htbp]
\centering
\begin{tabular}{rccc}
 & \textbf{Self-connectivity} & \textbf{Scattering} & \textbf{Homogeneity} \\ \hline \\[-0.25cm]
\textbf{Basal}& 0.57 & 1.56 & 0.77 \\[0.12cm]
\textbf{Mid} & 0.29 & 0.57 & 0.76 \\[0.12cm]
\textbf{Apical} & 0.28 & 0.73 & 0.75 \\[0.05cm] \hline  \end{tabular}
\caption{Quantitative indices, averaged from five repetitions with different seeds, of a
colored Mapper graph of the study dataset for a comparison between heart sites (basal, mid, or apical), obtained with percentages of change of three hemodynamic variables (LVP, DPDT+, ABF), \textbf{with minimum rising rate of left-ventricular peak pressure (DPDT--) eliminated from the analyses.}}
\label{table:BMA_APP}
\end{table}

\begin{table}[htbp]
\centering
\begin{tabular}{lccc}
& \textbf{Basal vs apical} & \;\textbf{Basal vs mid} & \textbf{Apical vs mid} \\ \hline \\[-0.25cm]
\textbf{\boldmath$p$-value} & $0.0327 ^{*}$ & \;$ 0.0069^{**}$ & $0.2935$
\\[0.1cm] \hline
\end{tabular}
\caption{Significance values of comparison tests for self-connectivity indices in Table~\ref{table:BMA_APP}. Medians of $p$-values after five repetitions are shown (*: $p<0.05$; **: $p<0.01$; ***: $p<0.001$).
Variables: LVP, DPDT+, ABF.}
\label{table:BA-BM-AM_APP}
\end{table}

\begin{table}[htbp]
\centering
\begin{tabular}{rccc}
 & \textbf{Self-connectivity} & \textbf{Scattering} & \textbf{Homogeneity} \\ \hline \\[-0.25cm]
\textbf{Endocardial} & 0.45  & 1.27   & 0.78 \\[0.12cm] 
\textbf{Epicardial} & 0.31 & 0.73   & 0.72  
\\[0.05cm]
\hline 
\end{tabular}
\caption{Quantitative indices, averaged from five repetitions with different seeds, of a
colored Mapper graph of the study dataset for a comparison between endocardial and epicardial stimulation,
obtained with percentages of change of three hemodynamic variables (LVP, DPDT+, ABF), \textbf{with minimum rising rate of left-ventricular peak pressure (DPDT--) eliminated from the analyses.}}
\label{table:EE_APP}
\end{table}

\begin{table}[htbp]
\centering
\begin{tabular}{lc}
& \textbf{Endocardial vs epicardial} \\ \hline \\[-0.25cm]
\textbf{\boldmath$p$-value} & $0.0397^*$ \\[0.1cm] \hline
\end{tabular}
\caption{Significance $p$-value of a comparison test for self-connectivity indices in Table~\ref{table:EE_APP}. Median of $p$-values after five repetitions is given.
Variables: LVP, DPDT+, ABF.}
\label{table:EEP_APP}
\end{table}

\begin{table}[htbp]
\centering
\begin{tabular}{rlllllll}
 & \textbf{ALP} & \textbf{AL} & \textbf{AP} & \textbf{LP} & \textbf{A} & 
 \textbf{L} &
 \textbf{P} \\ \hline \\[-0.2cm]
\textbf{With apical}\;\;\; & $0.0700$ & $0.0605$ & $0.1654$ & $0.1271$ & $0.1914$ & $0.0125^*$ & $0.1449$ \\[0.1cm]
\textbf{Without apical}\;\;\; & $0.2685$ & $0.0222^*$ & $0.3874$ & $0.1280$ & $0.0704$ & $0.0216^*$ & $0.1239$ \\[0.05cm] \hline
\end{tabular}
\caption{Comparison between endocardial and epicardial sites at each combination of regions (A: anterior; L: lateral; P: posterior) \textbf{with minimum rising rate of left-ventricular peak pressure (DPDT--) eliminated from the analyses,} Medians of $p$-values are shown for self-connectivity indices after five repetitions of the Mapper calculation (*: $p<0.05$; **: $p<0.01$; ***: $p<0.001$).
Variables: LVP, DPDT+, ABF. First row: including apical data in all columns; second row: excluding apical data.}
\label{table:ALP_APP}
\end{table}

\begin{table}[htbp]
\centering
\begin{tabular}{rll}
& \textbf{Epicardial} & \textbf{Endocardial} \\ \hline \\[-0.25cm]
\textbf{Basal vs mid}    & 
\;\;$0.0098^{**}$             & \;\;$0.0054^{**}$              \\[0.12cm]
\textbf{Basal vs apical} &      
\;\;$0.3036$             & \;\;$0.0350^*$ \\[0.05cm] \hline      
\end{tabular}
\caption{Significance $p$-values of comparison tests for self-connectivity indices of heart sites separated into epicardial and endocardial stimulation \textbf{with minimum rising rate of left-ventricular peak pressure (DPDT--) eliminated from the analyses.}
Variables: LVP, DPDT+, ABF.}
\label{table:BMBAEpiEndo_APP}
\end{table}

\begin{table}[htbp]
\centering
\begin{tabular}{rlllllll}
 & \textbf{ALP} & \textbf{AL} & \textbf{AP} & \textbf{LP} & \textbf{A} & 
 \textbf{L} &
 \textbf{P} \\ \hline \\[-0.2cm]

\textbf{All} & 
$0.0139^*$ &
$0.0039^{**}$ &
$0.0260^*$ &
$0.3535$ &
$0.3051$ &
$0.0003^{***}$ &
$0.2106$ \\[-0.3cm] \\

\textbf{w/o FIS06} & 
$0.0090^{**}$ &
$0.0066^{**}$ &
$0.1683$ &
$0.0662$ &
$0.0652$ &
$0.1239$ &
$0.3443$ \\[-0.3cm] \\

\textbf{w/o FIS13} & 
$0.0765$ & 
$0.3068$ & 
$0.1429$ & 
$0.1932$ & 
$0.4493$ & 
$0.1129$ & 
$0.3371$ \\[-0.3cm] \\

\textbf{w/o FIS14} & 
$0.0182^*$ & 
$0.2435$ & 
$0.0255^*$ & 
$0.0196^*$ & 
$0.0953$ & 
$0.0267^*$ & 
$0.1121$ \\[-0.3cm] \\

\textbf{w/o FIS15} & 
$0.2683$ & 
$0.2687$ & 
$0.0616$ & 
$0.0874$ & 
$0.2493$ & 
$0.1953$ & 
$0.1639$ \\[-0.3cm] \\

\textbf{w/o FIS18} & 
$0.0120^*$ & 
$0.3314$ & 
$0.0223^*$ & 
$0.2256$ & 
$0.0636$ & 
$0.0643$ & 
$0.3682$ \\[-0.3cm] \\

\textbf{w/o FIS19} & 
$0.0470^*$ & 
$0.0077^{**}$ & 
$0.0763$ & 
$0.1955$ & 
$0.2548$ & 
$0.0061^{**}$ & 
$0.2649$ \\[0.05cm] \hline

\end{tabular}
\caption{Comparison between endocardial and epicardial sites at each combination of regions (A: anterior; L: lateral; P: posterior), excluding one individual at a time (coded FIS06 to FIS19). The first row includes all six study subjects.
Medians of $p$-values for self-connectivity indices after five repetitions of the Mapper calculation are given. Significance of $p$-values: ${}^*<0.05$, ${}^{**}<0.01$, ${}^{***}<0.001$. Variables: LVP, DPDT+, DPDT--, ABF. Apical data are included in all columns.}
\label{table:excl}
\end{table}

\begin{table}[htbp]
\centering
\begin{tabular}{rlllllll}
 & \textbf{ALP} & \textbf{AL} & \textbf{AP} & \textbf{LP} & \textbf{A} & 
 \textbf{L} &
 \textbf{P} \\ \hline \\[-0.2cm]

\textbf{All} & 
$0.0700$ & 
$0.0605$&
$0.1654$ & 
$0.1271$ & 
$0.1914$ & 
$0.0125^*$ & 
$0.1449$ \\[-0.3cm] \\

\textbf{w/o FIS06} & 
$0.0338^*$ & 
$0.0215^*$ & 
$0.1847$ & 
$0.0201^*$ & 
$0.0382^*$ & 
$0.0053^{**}$ & 
$0.1925$
 \\[-0.3cm] \\

\textbf{w/o FIS13} & 
$0.0218^*$ & 
$0.0328^*$ & 
$0.0099^{**}$ & 
$0.1045$ & 
$0.1451$ & 
$0.2389$ & 
$0.1065$ \\[-0.3cm] \\

\textbf{w/o FIS14} & 
$0.1040$ & 
$0.0235^*$ & 
$0.0225^*$ & 
$0.1600$ & 
$0.2152$ & 
$0.1239$ & 
$0.0485^*$ \\[-0.3cm] \\

\textbf{w/o FIS15} & 
$0.1903$ & 
$0.2551$ & 
$0.0901$ & 
$0.3650$ & 
$0.0181^*$ & 
$0.4077$ & 
$0.0808$ \\[-0.3cm] \\

\textbf{w/o FIS18} & 
$0.0932$ & 
$0.2063$ & 
$0.3423$ & 
$0.3988$ & 
$0.1958$ & 
$0.1226$ & 
$0.2003$ \\[-0.3cm] \\

\textbf{w/o FIS19} & 
$0.0741$ & 
$0.0519$ & 
$0.1395$ & 
$0.1517$ & 
$0.1935$ & 
$0.0531$ & 
$0.2806$ \\[0.05cm] \hline

\end{tabular}
\caption{Comparison between endocardial and epicardial sites at each combination of regions (A: anterior; L: lateral; P: posterior) 
\textbf{with minimum rising rate of left-ventricular peak pressure (DPDT--) eliminated from the analyses} and
excluding one individual at a time (coded FIS06 to FIS19). The first row includes all six study subjects.
Medians of $p$-values for self-connectivity indices after five repetitions of the Mapper calculation are given. Significance of $p$-values: ${}^*<0.05$, ${}^{**}<0.01$, ${}^{***}<0.001$. Variables: LVP, DPDT+, ABF. Apical data are included in all columns.}
\label{table:excl_noDPDT}
\end{table}

\end{document}